\documentstyle[12pt,amssymb,righttag]{amsart}
\newtheorem{thm}{Theorem}[section]

\newtheorem{prop}[thm]{Proposition}
\newtheorem{defin}[thm]{Definition}

\theoremstyle{remark}

\font\smc=cmcsc10 at 12pt
\font\eightsmc=cmcsc8 at 12pt
\def\R{{\Bbb R}}\def\Z{{\Bbb Z}}\def\C{{\Bbb C}}
\def\re{\Re e\,}
\let\o=\operatorname
\def\oskip{\par\vbox to4mm{\vfill}\par}
\def\doublecheck#1{{\kern2.1pt\hbox{$\check{\kern-2.1pt\hbox{$\check #1$}}$}}}
\begin{document}\baselineskip=15pt
\pagestyle{plain}
\rightline{February 24, 1998}
\rightline{physics/9802044}
\rightline{to appear in \sl Lett. Math. Phys.}
\oskip\oskip
{\parskip=0pt
\begin{center}
{\bf MIRROR SYMMETRY ON  K3 SURFACES}
\par\medskip
{\bf AS A  HYPERK\"AHLER ROTATION}
\par\bigskip\bigskip
{\smc Ugo Bruzzo}\thinspace \S\ddag \ \ and \
{\smc Guido Sanguinetti \thinspace \P\S}
\par\bigskip
{\S\thinspace Scuola Internazionale Superiore di Studi Avanzati,}
\par {Via Beirut 2-4, 34014 Trieste, Italy}
\par\smallskip
{\ddag\thinspace Dipartimento di Matematica, Universit\`a degli Studi}
\par{di Genova, Via Dodecaneso 35, 16146 Genova, Italy}
\par\smallskip
{\P\thinspace Dipartimento di Scienze Fisiche, Universit\`a degli Studi}
\par{di Genova, Via Dodecaneso 33, 16146 Genova, Italy}
\par\smallskip
{E-mail: {\tt bruzzo@@sissa.it}, {\tt sanguine@@sissa.it}}
\end{center}
\par\oskip\oskip
{\footnotesize\par\leftskip=40pt\rightskip=40pt
{\eightsmc Abstract.}  We show that under the hypotheses
of \cite{SYZ}, a   mirror partner
of a K3 surface $X$ with a  fibration in special Lagrangian tori can be obtained
by rotating the complex structure of $X$ within its hyperk\"ahler family of
complex structures. Furthermore, the same hypotheses force the B-field to
vanish.\par} }

\oskip\oskip
\section{Introduction}
According to the proposal of Strominger, Yau and Zaslow \cite{SYZ}, the mirror
partner of a K3 surface $X$ admitting a  fibration in special Lagrangian tori
should be identified with the moduli space of such fibrations (cf.~also
\cite{H}). In more precise terms, the mirror partner $\check X$  should be
identified with  a suitable compactification of the relative Jacobian of $X'$,
where $X'$ is an elliptic K3 surface obtained
by rotating the complex structure of $X$ within its hyperk\"ahler family
of complex structures.

Morrison \cite{M} suggested that such a compactification is provided by the
moduli space of torsion sheaves of degree zero and pure dimension one supported
by the fibers of $X'$. (It should be noted that whenever the fibration
$X'\to{\Bbb P}^1$ admits a holomorphic section, as it is usually assumed in the
physical literature, the complex manifolds $X'$ and $\check X$ turn out to be
isomorphic). In \cite{BBHM} Morrison's suggestion was implemented, and it was
shown that the  relative Fourier-Mukai transform
defined by the Poincar\'e sheaf on the fiber product $X'\times_{{\Bbb
P}^1}\check X$  enjoys some good properties related to mirror symmetry; e.g., it
correctly maps D-branes in $X$ to D-branes in $\check X$, preserves the masses
of the BPS states, etc. (The fact that the Fourier-Mukai transform
might describe some aspects of mirror symmetry was already suggested in
\cite{DM}.)

It remains to check that $\check X$ is actually a mirror of $X$ in the sense of
Dolgachev and Gross-Wilson, cf.~\cite{D,GW,G}.
In this note we show that this
is indeed the case. Roughly speaking, we prove that whenever $X$ admits a
fibration in special Lagrangian tori with a section, and also admits an
elliptic mirror $\check X$ with a section,\footnote{These are the same
assumptions made in \cite{SYZ} on physical grounds.} then the complex structure
of
$\check X$ is obtained by that of $X$ by redefining the B-field and then
performing a hyperk\"ahler rotation. A more precise statement is as follows. Let
$M$ be a primitive sublattice of the standard 2-cohomology lattice of a K3
surface, and denote by
${\bold K}_M$ the moduli space of pairs $(X,j)$, where $X$ is a K3 surface, and
$j\colon M \to \o{Pic}(X)$ is a primitive lattice embedding. Let $T=M^\perp$.
We assume that $T$ contains a $U(1)$  lattice $P$; this means that the generic
K3 surface $X$ in ${\bold K}_M$, possibly after a rotation of its complex
structure within its hyperk\"ahler family, admits a fibration in special
Lagrangian tori with a section.  After setting
$\check M=T/P$, we assume that the generic K3 surface in ${\bold K}_{\check M}$
is elliptic and has a section. These hypotheses force the B-field to be an
integral class. Then, by setting to zero this class (as it seems to be
suggested by the physics, since in string theory the B-field is a class in
$H^2(X,\R/\Z)$), and rotating the complex structure of $X$ within its
hyperk\"ahler family of complex structures, we associate to
$X\in {\bold
K}_M$  a K3 surface $\check X$ in
${\bold K}_{\check M}$ such that $\o{Pic}(\check X)\simeq\check M$.

\oskip\section{Special Lagrangian fibrations and mirror K3
surfaces\label{prelim}}
We collect here, basically relying on \cite{HL,Mc,D,GW}, some basic definitions
and constructions about mirror families of K3 surfaces.
\par\smallskip {\it Special Lagrangian submanifolds.} Let $X$ be an
$n$-dimensional K\"ahler manifold with K\"ahler form
$\omega$, and suppose that on $X$ there is a nowhere vanishing holomorphic
$n$-form $\Omega$. One says that a real $n$-dimensional submanifold $\iota\colon
Y\hookrightarrow X$ is {\it special Lagrangian} if $\iota^\ast\omega=0$, and
$\Omega$ can be chosen so that the form  $\iota^\ast\re\Omega$ coincides with
the volume form of $Y$. The moduli space of deformations of $Y$ through special
Lagrangian submanifolds was described in \cite{Mc}.

Let $n=2$,  assume that $X$ is hyperk\"ahler with Riemannian metric $g$, and
choose basic complex structures
$I$, $J$, and $K$. These generate an $S^2$ of complex structures compatible
with the Riemannian metric of $X$, which we shall call the {\it hyperk\"ahler
family} of complex structures of $X$.

Denote by $\omega_I$, $\omega_J$ and $\omega_K$ the K\"ahler forms
corresponding to the complex structures $I$, $J$ and $K$. The 2-form
$\Omega_I=\omega_J+i\,\omega_K$ never vanishes, and is holomorphic with respect
to $I$. Thus, submanifolds of $X$ that are special Lagrangian with respect to
$I$, are holomorphic with respect to $J$ (this is a consequence of Wirtinger's
theorem, cf.~\cite{HL}). If
$X$ is a complex K3 surface that admits a foliation by special Lagrangian
2-tori (in the complex structure
$I$), then in the complex structure $J$ it is an elliptic surface, $p\colon
X'\to{\Bbb P}^1$. If one wants $X$ to be compact then one must allow the
fibration $p\colon X'\to{\Bbb P}^1$ to have some singular fibers, cf.~\cite{K}.
\par\smallskip
{\it Mirror families of K3 surfaces \cite{D}.}
Let $L$ denote the lattice over $\Z$
$$L = U(1) \perp  U(1) \perp  U(1) \perp E_8 \perp E_8$$
(by ``lattice over $\Z$'' we mean as usual a free finitely generated $\Z$-module
equipped with a symmetric $\Z$-valued quadratic form). If $X$ is a K3 surface,
the group $H^2(X,\Z)$ equipped with the cohomology intersection pairing is a
lattice isomorphic to $L$.

If $M$ is an even nondegenerate lattice of signature $(1,t)$, a {\it
$M$-polarized K3 surface} is a pair $(X,j)$, where
$X$ is a K3 surface and $j\colon M\to\o{Pic}(X) $ is a primitive lattice
embedding. One can define a coarse moduli space ${\bold K}_M$ of  $M$-polarized
K3 surfaces; this is a quasi-projective algebraic variety of  dimension
$19-t$, and may be obtained by taking a quotient of the space
$$D_M =\left\{\C \Omega \in {\Bbb P}(M^\perp\otimes \C )\,\vert\,\Omega
\cdot\Omega =0, \Omega \cdot \bar \Omega >0\right\}$$ by a
discrete group $\Gamma_M$ (which is basically the group of isometries of $L$
that fix all elements of $M$) \cite{D}.

A basic notion to introduce the {\it mirror moduli space} to ${\bold K}_M$ is
that of {\it admissible $m$-vector}. We shall  consider here only the case
$m=1$. Let us pick a primitive sublattice $M$ of $L$ of signature $(1,t)$.

\begin{defin} A 1-admissible vector $E\in M^\perp$ is an isotropic vector in
$M^\perp$ such that there exists another isotropic vector $E'\in M^\perp$ with
$E\cdot E'=1$.
\end{defin}

After setting
$$\check M = E^\perp/\Z E$$ one easily shows that there is an orthogonal
decomposition
$M^\perp=P\oplus\check M$, where $P$ is the hyperbolic lattice generated by $E$
and $E'$. The orthogonal of $E$  is taken here in $M^\perp $. The {\it mirror
moduli space} to ${\bold K}_M$ is the space ${\bold K}_{\check M}$. Of course
one has
$$\o{dim}{\bold K}_M+\o{dim}{\bold K}_{\check M}=20\,.$$ The operation of
taking the ``mirror moduli space'' is a duality, i.e.~$\doublecheck M\simeq M$
(this works so because we consider the case of a 1-admissible vector, and is no
longer true for $m>1$).

\smallskip {\it The interplay between special Lagrangian fibrations and mirror
K3 surfaces.} Let again  $M$ be an even nondegenerate lattice of signature
$(1,t)$, and suppose that $X$ is K3 surface such that $\o{Pic}(X)\simeq M$.
The  transcendental lattice $T$ (the orthogonal complement of
$\o{Pic}(X)$  in  $H^2(X,\Z)$) is an even lattice of signature
$(2, 19-t)$.  Let $\Omega =x+i\,y$ be a nowhere vanishing, global holomorphic
two-form on $X$. Being orthogonal to all  algebraic classes, the cohomology
class of $\Omega$  spans a space-like 2-plane in $T\otimes
\R$. The moduli space of K3 such that $\o{Pic}(X)\simeq M$  is parametrized by
the periods, whose real and imaginary parts  are given by intersection with
$x$ and $y$, respectively. Indeed, one should  recall  that if we fix a basis of
the cohomology lattice $H^2(X,\Z)$ given by integral cycles $\alpha_i$,
$i=1,\dots,22$, every complex structure on $X$ is uniquely determined, via
Torelli's theorem, by the complex valued matrix whose entries $\varpi_i$ are
given by the intersections of the cycles
$\alpha_i$ with the class of the holomorphic two-form $\Omega $, i.e.~$\varpi_i
=\alpha _i\cdot \Omega $.  This shows that generically neither $x$ nor $y$ are
integral classes in the cohomology  ring. However, if we make the further
request that there is a 1-admissible vector in $T$, and make some choices, one
of the two classes is forced to be integral.

We recall now a result from \cite{GW} (although in a slightly weaker form).
\begin{prop}
There exists in $T$ a 1-admissible vector if and only if there is a
complex structure on $X$ such that $X$ has a special Lagrangian
fibration with a section.
\end{prop}
So we consider on $X$ a complex structure satisfying this property (it follows
from \cite{GW} that, if we fix a hyperk\"ahler metric on $X$, this complex
structure belongs to the
same hyperk\"ahler family as the one we started from). As a direct
consequence we have
\begin{prop} If there exists a 1-admissible vector in $T$ one can perform a
hyperk\"ahler rotation of the complex structure and choose a nowhere vanishing
two-form
$\Omega $, holomorphic in the new complex structure, whose real part $\re\Omega$
is integral.
\end{prop}
\begin{pf}
By Proposition 1.3 of \cite{GW} the existence of a 1-admissible vector implies
the existence on $X$ of a special Lagrangian fibration with a section. On the
other hand  by \cite{HL}  what is special Lagrangian in a complex structure  is
holomorphic in the complex structure in which the K\"ahler form is given by
$\re\Omega $. Thus in this complex structure the Picard group is nontrivial,
which implies that the surface is algebraic, i.e.~$\re\Omega$ is integral.
\end{pf}

\oskip\section{The construction}
We introduce now a moduli space $\tilde{\bold K}_M$ parametrizing
$M$-polarized K3 surfaces together with of a 1-admissible vector
in $T=M^\perp$. The generic K3 surface $X$ in $\tilde{\bold K}_M$
admits a fibration in special Lagrangian tori
with a section; the primitive $U(1)$
sublattice $P$ of the transcendental lattice $T$ associated with the
1-admissible vector is generated by the class of the fiber and the class of the
section. We fix a marking\footnote{Since we are fixing
a marking of $X$ in the following we shall often confuse the lattices
$H^2(X,\Z)$ and $L$.} of $X$, i.e., a lattice isomorphism
$\psi\colon H^2(X,\Z)\to L$. We have an isomorphism
$$L\simeq M \oplus P \oplus \check M,$$
where  $\check M = T/P$.
The fact that $\doublecheck M\simeq M$ implies that the moduli spaces
$\tilde{\bold K}_M$ and $\tilde{\bold K}_{\check M}$ are isomorphic.
Generically, we may assume that  $M\simeq\psi(\o{Pic}(X)$).

 One easily shows that the following
assumptions are generically equivalent to each other (where ``generically''
means that this holds true for $X$ in a dense open subset of $\tilde{\bold
K}_M$):

(i) The lattice $\check M$ contains a primitive $U(1)$ sublattice $P'$.

(ii) The generic K3 surface in the mirror moduli space ${\bold
K}_{\check M}$ is an elliptic fibration with a section.

(iii) $X$ carries two fibrations in special
Lagrangian tori admitting a section, in such a way that the corresponding
$U(1)$ lattices $P$, $P'$ are orthogonal.\footnote{Then one shows that
the direct sum $P\oplus P'$ is an orthogonal summand of $T$.}

The two $U(1)$ lattices $P$ and $P'$ are
interchanged by an isometry of $L$. Thus, the operation of exchanging them
has no effect on the moduli space ${\bold K}_M$ (although it does on $D_M$).

We shall assume one of these equivalent conditions.
The form (ii) of the second condition shows that we are working exactly under
the same assumptions that in \cite{SYZ} are advocated on physical grounds.

In the complex structure of $X$   we have fixed at the outset we  have the
K\"ahler form $\omega $ and the holomorphic two-form $\Omega = x+i\,y$, with $x$
an integral class. Condition (iii) means that $P'$ is calibrated by $x$.
If we perform a rotation around the $y$ axis, mapping the pair $(\omega,x)$ to
$(x,-\omega)$, we  still obtain an algebraic K3 surface $X'$ whose Picard group
contains $P'$ \cite{GW}.

Now we want to show that the K\"ahler class of $X'$ is a space-like vector
contained in the hyperbolic lattice $P'$. We remind here that the explicit
mirror map in \cite{D} and
\cite{GW}  is given in terms of a choice of a hyperbolic sublattice of the
transcendental  lattice. Let $D_M$ be defined as in Section \ref{prelim}, and
let
$$T_M =\left\{B+i\,\omega \in M\otimes \C \,\vert\,\omega
\cdot\omega>0\right\}=M\times V(M)^+\,.$$
Here $V(M)^+$ is the component of the positive cone in $M\otimes \R$
that contains the K\"ahler form of $X$. The space
$T_M$ can be regarded as a (covering of the) moduli space of ``
complexified K\"ahler structures''
on $X$. Let $M'=T/P'\simeq \check M$. By \cite{GW} Proposition 1.1,
the mirror map is an isomorphism
$$\phi \colon T_{M'}\to D_M\,,$$
$$\phi (\check B+i\check \omega )=\check B + E'+\tfrac12(\check \omega \cdot
\check \omega -\check B\cdot\check B)E+i\,(\check \omega -(\check \omega
\cdot\check B)E)\,.
$$ Here $E$ and $E'$ are the two isotropic generators of the
$U(1)$ lattice $P'$, while $\check B$ is what the physicists call the B-field.
Our holomorphic two-form $\Omega $ is of course of the form
$\phi (\check B+i\check \omega)$ for suitable $\check B$ and $\check \omega$,
since $\phi $ is an isomorphism. The  K\"ahler class of $X'$ is given by
$$x=\re\Omega = \check B + E' +\tfrac12(\check\omega\cdot \check\omega -
\check B
\cdot\check B)E$$
and the new global holomorphic two-form is $-\omega +iy$. Since $\check B$ is
orthogonal to $E$ and $E'$, it is an integral class.

However,
the Picard lattice of the K3 surface $X'$ is generically not isomorphic
to $\check M$. A better
choice is suggested by the physics. Indeed in most string theory models
the B-field is regarded as a Chern-Simons term, namely, as a class
in $H^2(X,\R/\Z)$; so, if we consider the projection
$ \lambda\colon H^2(X,\R) \to H^2(X,\R/\Z)$,
the relevant moduli space should be
$$\tilde T_{M'} = \lambda(M'\otimes\R)\times V(M')^+$$
instead of $T_{M'}$.
To take this suggestion into account we set $\check B=0$. Since
$y=\check\omega-(\check\omega\cdot\check B)E$, this changes the complex
structure in $X'$. Moreover, $x$ lies now in $P'$.

So, let us now consider the intersection of $P\otimes \R$ with the spacelike
two-plane $\langle\Omega\rangle $ spanned by $\Omega $. This cannot be trivial,
since
$P$ is hyperbolic and $T\otimes \R$ is of signature $(2,19-t)$. So we have a
real space-like class in $P\otimes \R \cap \langle\Omega\rangle $ that is
orthogonal to $x$ by construction and thus must be equal (up to a scalar
factor) to $y$. But then, in the complex structure in which the K\"ahler form
is given by $x$, all the cycles of $\check M$ are orthogonal to the new
holomorphic two-form, given by
$\omega +iy$, and therefore are algebraic.
(Notice that the class $y$ is not integral.)

\oskip\section {Conclusions}
A first conclusion we may draw is that the hypotheses of \cite{SYZ}
force the B-field to be integral, namely, to be zero as a class
$\check B\in H^2(X,\R/\Z)$. Moreover,
starting from a K3 surface $X$ in $\tilde{\bold K}_M$,
the construction in the previous section
singles out a point in the variety
$\tilde{\bold K}_{\check M}$; so we have
established a map
$$\mu\colon\tilde{\bold K}_{M}\to \tilde{\bold K}_{\check M}$$
which is bijective by construction, and deserves to be the called {\it the
mirror map}. This map consists in setting $\check B$ to zero  (as a
class in
$H^2(X,\Z)$) and then performing a hyperk\"ahler rotation.

If we do not set $\check B$ to zero, we obtain a family
of K3 surfaces, labelled by the possible values of  $\check B\in M'\simeq\check
M$. Its counterpart under mirror symmetry is a family of  K3 surfaces
labelled by $M$. The two families are related by a hyperk\"ahler rotation.

\medskip{\bf Acknowledgements.} We thank C. Bartocci, I. Dolgachev and
D. Her\-n\'an\-dez Rui\-p\'e\-rez for useful comments and discussions. This work
was partly supported by Ministero dell'Universit\`a e della Ricerca Scientifica
e Tecnologica through the research project ``Geometria reale e complessa.''

\oskip

\end{document}